\documentclass[aps,pra,amsmath,amssymb,reprint,superscriptaddress,longbibliography]{revtex4-2}
\usepackage{xcolor}
\usepackage{graphicx}
\usepackage{dcolumn}
\usepackage{bm}
\usepackage{braket}
\usepackage{verbatim} 
\usepackage[english]{babel}
\usepackage{xfrac}
\usepackage{soul}
\usepackage{comment}
\usepackage{braket}
\usepackage[explicit]{titlesec}

\begin{document}


\title{Energy Transport Among Highly-Polarized Atoms}

\author{Catherine D. Opsahl}
\affiliation{Department of Physics, Bryn Mawr College, Bryn Mawr, PA 19010.}

\author{Yuan Jiang}
\affiliation{Department of Physics, Bryn Mawr College, Bryn Mawr, PA 19010.}

\author{Samantha A. Grubb}
\affiliation{Department of Physics and Astronomy, Ursinus College, Collegeville, PA 19426.}

\author{Alan T. Okinaka}
\affiliation{Department of Physics and Astronomy, Ursinus College, Collegeville, PA 19426.}

\author{Nicolaus A. Chlanda}
\affiliation{Department of Physics and Astronomy, Ursinus College, Collegeville, PA 19426.}

\author{Hannah S. Conley}
\affiliation{Department of Physics and Astronomy, Ursinus College, Collegeville, PA 19426.}

\author{Aidan D. Kirk}
\affiliation{Department of Physics and Astronomy, Ursinus College, Collegeville, PA 19426.}

\author{Sarah E. Spielman}
\affiliation{Department of Physics, Bryn Mawr College, Bryn Mawr, PA 19010.}

\author{Thomas J. Carroll}
\affiliation{Department of Physics and Astronomy, Ursinus College, Collegeville, PA 19426.}

\author{Michael W. Noel}%
\affiliation{Department of Physics, Bryn Mawr College, Bryn Mawr, PA 19010.}




\date{\today}

\begin{abstract}
We measure the transport of energy among the internal states of ultracold rubidium Rydberg  atoms coupled by dipole-dipole exchange. In a magneto-optical trap, a static electric field of a few V/cm shifts the energy levels of the atoms. For a particular principal quantum number, $n$, the angular momentum eigenstates $\ell > 4$ are nearly degenerate at zero electric field.  At nonzero field, a manifold of equally spaced clusters form a ladder with each rung consisting of a set of closely spaced $m$ energy eigenstates. We excite Rydberg atoms to energy levels near the center of the manifold and allow them to exchange energy via resonant dipole-dipole interactions. We measure the time evolution as energy spreads away from the center of the manifold, which reveals that the system may fail to thermalize for long interaction times. A computational model that includes only a few essential features of the system qualitatively agrees with this result. 

\end{abstract}


\maketitle

\section{Introduction}
Understanding transport dynamics in quantum systems is essential for answering fundamental questions surrounding thermodynamics and quantum many-body phenomena. Transport physics underlie a myriad of effects, ranging from superconductivity and thermal exchange in solid-state systems, to energy transfer in photosynthetic systems~\cite{engel_EvidenceWavelikeEnergy_2007}, or even that of many-body localization, in which the suppression of transport prevents thermalization~\cite{pal_ManybodyLocalizationPhase_2010}.

Ultracold atoms, which are highly controllable and easily isolated, have proven to be viable quantum simulators of solid-state physics, exploring conductivity and thermoelectric effects~\cite{brantut_ConductionUltracoldFermions_2012,brantut_ThermoelectricHeatEngine_2013, brown_BadMetallicTransport_2019}, in addition to building hydrodynamic frameworks of quantum systems~\cite{schemmer_GeneralizedHydrodynamicsAtom_2019, malvania_GeneralizedHydrodynamicsStrongly_2021, schuttelkopf_CharacterizingTransportQuantum_2026}. The sensitivity of cold atoms to external perturbations is also ideal for realizing out-of-equilibrium many-body dynamics~\cite{schneider_FermionicTransportOutofequilibrium_2012, trotzky_ProbingRelaxationEquilibrium_2012a, ronzheimer_ExpansionDynamicsInteracting_2013, scherg_NonequilibriumMassTransport_2018} and nonthermal behaviors~\cite{kohlert_ObservationManyBodyLocalization_2019, schreiber_ObservationManyBodyLocalization_2015a, smith_ManybodyLocalizationQuantum_2016,  choi_ExploringManybodyLocalization_2016a, guardado-sanchez_SubdiffusionHeatTransport_2020}. Optical lattices offer even more precise control of interactions, enabling clear observations of transport of fermions and bosons~\cite{strohmaier_InteractionControlledTransportUltracold_2007, schneider_FermionicTransportOutofequilibrium_2012,trotzky_ProbingRelaxationEquilibrium_2012a, ronzheimer_ExpansionDynamicsInteracting_2013, choi_ExploringManybodyLocalization_2016a, ronzheimer_ExpansionDynamicsInteracting_2013,scherg_NonequilibriumMassTransport_2018,guardado-sanchez_SubdiffusionHeatTransport_2020}, excitations~\cite{ramm_EnergyTransportTrapped_2014,barredo_CoherentExcitationTransfer_2015b}, 
and of anomalous spin dynamics in 2D~\cite{hild_FarfromEquilibriumSpinTransport_2014, nichols_SpinTransportMott_2019, joshi_ObservingEmergentHydrodynamics_2022}. G{\"u}nter~\textit{et al.} and Fahey~\textit{et al.} have imaged samples of Rydberg atoms directly to measure the spatial redistribution of electronic energy states arising from dipole-dipole energy exchange~\cite{gunter_ObservingDynamicsDipoleMediated_2013, fahey_ImagingDipoledipoleEnergy_2015}.

Another approach that has gained recent attention is to encode internal states as additional degrees of freedom, or \textit{synthetic dimensions}. They have been used to engineer higher-dimensional Hamiltonians using laser-driven transitions between adjacent states, and in the case of atomic systems, simulate hopping across an additional lattice dimension~\cite{boada_QuantumSimulationExtra_2012, celi_SyntheticGaugeFields_2014a}. Synthetic dimensions have been used to study topological states, phase transitions, and many-body dynamics~\cite{livi_SyntheticDimensionsSpinOrbit_2016,an_DiffusiveArrestedTransport_2017,sundar_StringsUltracoldMolecules_2019, kanungo_RealizingTopologicalEdge_2022,lu_WavepacketDynamicsLongrange_2024,lu_ProbingTopologicalPhase_2024,chen_StronglyInteractingRydberg_2024}, with many of these recent experiments constructed with Rydberg atoms~\cite{boada_QuantumSimulationExtra_2012,kanungo_RealizingTopologicalEdge_2022, lu_WavepacketDynamicsLongrange_2024, lu_ProbingTopologicalPhase_2024, chen_StronglyInteractingRydberg_2024}. Recently, Kanungo~\textit{et al.} used ladders of $ns$ and $np$ Rydberg levels coupled with millimeter waves as a synthetic dimension to probe band structure~\cite{kanungo_RealizingTopologicalEdge_2022}.

We introduce a new experimental platform, in which we measure the energy transport in nearly harmonic ladders of Stark energy levels in ultracold Rydberg atoms. The energy exchange is mediated by resonant dipole-dipole interactions among the amorphous cloud of rubidium Rydberg atoms in a magneto-optical trap (MOT). The sensitivity of Rydberg atoms to external electric and magnetic fields offers some control over the parameters of the ladder, making this a promising model system.

\begin{figure}
 \includegraphics{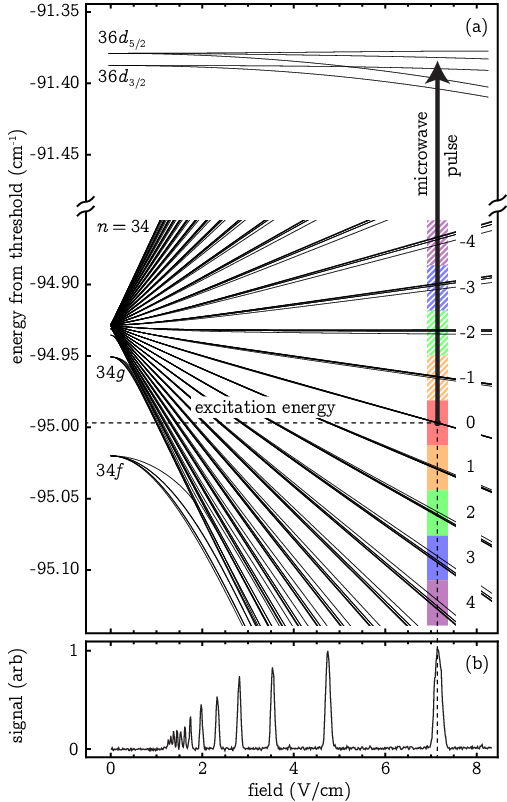}%
\caption{\label{fig:starkmap}(a) Stark map showing the $|m_j|=1/2, 3/2$, and 5/2 states of the $n=34$ manifold and the $36d$ states. These manifold states are organized into clusters of energy levels with nearly harmonic spacing. The initially excited manifold cluster is highlighted in red and labeled 0. During an interaction time~$\le 1\ \mu s$, resonant dipole-dipole interactions transfer population to clusters above and below the initial cluster, which are highlighted in different colors and numerically labeled relative to the initial cluster. A microwave pulse is scanned over frequency to transfer the population of each manifold cluster to the $36d$ state, where it can be resolved with state selective field ionization. (b) An electric field scan at a particular wavelength of the Rydberg excitation laser, showing each manifold cluster intersected by the horizontal dashed line in (a).}
\end{figure}

When an atom is immersed in an electric field, the degeneracy among different angular momentum states is lifted.  In non-relativistic hydrogen, the state splittings are dominated by the linear Stark effect, which leads to a harmonic spacing.  While the $\ell$ degeneracy is lifted, the $m$ degeneracy is not.  Thus, each rung of the ladder consists of a set of even or odd $m$ states with the number of degenerate states increasing at the center of the manifold.  There are, however, some deviations from this in alkali atoms due to the complexity of the ion core.  The quantum defects lift the $\ell$ degeneracy at zero field leading to a quadratic Stark shift at low field.  Above $\ell=3$, the quantum defects are small so this quadratic shift quickly returns to a nearly linear shift as the field increases.  The shift of state energies with electric field near $n=34$ is shown in the Stark map of Fig.~\ref{fig:starkmap}(a).

Including fine structure and considering only the low $m_j$ states in rubidium, we can visualize the manifold as clusters of six energy levels each with $|m_j| = \frac{1}{2}, \frac{3}{2}, \frac{5}{2}$. For an initial cluster near the center of the manifold, dipole moments between clusters vary between 0 and 60~$ea_0$ whereas dipole moments within clusters can be much stronger, around $250$~$ea_0$. At our chosen field for this experiment, clusters are separated by about 0.8~GHz and the initial cluster is about 14~MHz wide. The closest pair of energy levels within the initial cluster are separated by about 0.4~MHz.

For a sample of atoms initially excited to a state near the middle of the Stark manifold, we might expect dipole-dipole interactions to drive the system into thermal equilibrium, with population spread uniformly across the manifold.  Instead, we find that a significant fraction of the population remains in the initial state, even at long interaction times. A simple model of this system confirms this behavior, pointing to the interplay between interactions within a cluster and across rungs of the ladder as the cause of the non-thermal dynamics.

\section{Experiment}

\begin{figure}
 \includegraphics{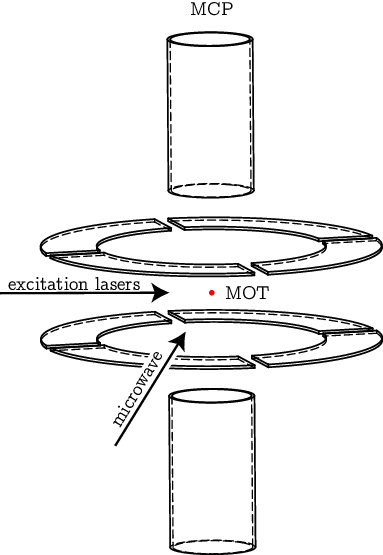}%
\caption{\label{fig:electrodes}Experimental geometry. A set of cylinders and segmented rings are used to provide the electric fields.  In this experiment, the top cylinder and both segmented rings remain grounded.  A steady voltage is applied to the bottom cylinder to set the electric field at the MOT, which is located at the center of the electrode configuration.  The field ionization pulse is also applied to the bottom cylinder.  Upon ionization, electrons are accelerated to the micro-channel plate (MCP) detector.  Rydberg excitation lasers and microwave fields enter through the side of the electrode configuration.}
\end{figure}

Our experiment, shown in Fig.~\ref{fig:electrodes}, uses $^{85}$Rb atoms that are cooled and confined in a magneto-optical trap.  With the trapping laser driving the $5s \rightarrow 5p$ cycling transition, we use two additional lasers to excite Rydberg atoms.  One laser operates at 776~nm to drive the $5p \rightarrow 5d$ transition and the other excites from $5d$ to states with $nf$ character at 1265~nm.  The 780~nm and 776~nm lasers each have a linewidth of roughly 100~kHz and are locked to the appropriate transitions using saturated absorption spectroscopy and Doppler free electromagnetically induced transparency respectively.  The frequency of the 1265~nm laser is locked to a Fabry-Perot cavity with a long term drift of a few MHz over the 24 hour period of a typical data run. The 776~nm and 1265~nm lasers are overlapped and focused through the MOT, creating a cylindrical volume of Rydberg excitation roughly 0.5~mm long with a diameter of 50~$\mu$m.

\begin{figure}
 \includegraphics{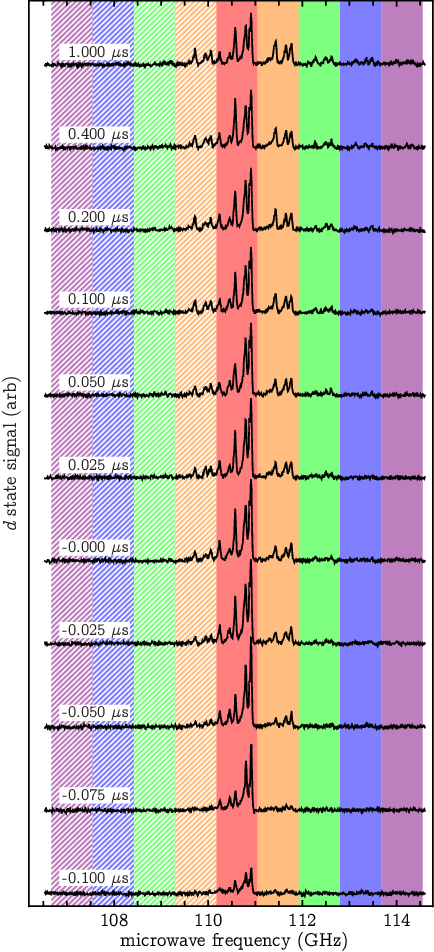}%
\caption{\label{fig:microwave}Field ionization signal of the $36d$ state as a function of microwave frequency, shown at different delays. Each shaded region corresponds to a cluster of manifold energy levels in Fig.~\ref{fig:starkmap}(a). Each peak within a shaded region is due to a different $36d_{j, |m_j|}$ sublevel. All five possible peaks are most easily visible in the central, red-shaded region corresponding to the initially excited cluster. With increasing delay, dipole-dipole interactions transfer population from the initial cluster to adjacent clusters. For 0~$\mu s$, the end of the Rydberg excitation pulse coincides with the start of the microwave pulse. }
\end{figure}

Figure~\ref{fig:starkmap}(a) shows a Stark map for the region of interest.  We must first find the desired Stark cluster.  This is done by locking the frequency of the 1265~nm laser to an energy just below the zero-field high-$\ell$ $n=34$ manifold.  We then measure the number of excited Rydberg atoms as the field is increased, as shown in Fig.~\ref{fig:starkmap}(b).  Each peak in this figure represents a cluster of Stark states tuning into resonance with the 1265~nm laser at a particular static electric field.  The zero field $f$ character is spread across the Stark manifold, even at low field, allowing excitation of a broad range of clusters. In this experiment we chose the cluster that tunes into resonance at 7.15~V/cm.  For this initial cluster, neighboring clusters are nearly harmonically spaced and the spacings of energy levels within each cluster is small.  We do not have sufficient excitation resolution to resolve the energy levels within a cluster, so our initial state is a mixture of all energy levels within this cluster.  Once the desired cluster is found, the static electric field remains fixed at this value.  

Upon excitation to the Stark cluster, atoms can immediately begin to exchange energy through dipole-dipole interactions, which spreads population to neighboring clusters.  To study this energy transport, we must measure the distribution of population across the Stark manifold.  Selective field ionization cannot resolve the signal from individual Stark energy levels or clusters, but it can resolve the $36d$ energy level from the $n=34$ manifold.  We therefore use a pulsed microwave field as shown in Fig.~\ref{fig:starkmap}(a) to drive population from the manifold states to the $36d$ state.  The microwave field is generated with an HP83623A sweep oscillator, whose output is sent through an active frequency doubler, followed by a passive frequency tripler.  The final microwave output is emitted from a wr10 horn into the vacuum system through a 6~inch viewport.  The horn is placed at an angle with respect to the applied static electric field so that transitions of $\Delta m = 0$ and $\Delta m = \pm 1$ can be driven.  The tuning range of this microwave system is 75 - 120 ~GHz.

\begin{figure}
 \includegraphics{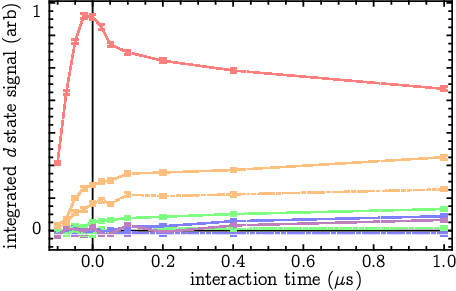}%
\caption{\label{fig:time} Integrated signal as a function of time for each of the highlighted manifold clusters in Figs.~\ref{fig:starkmap}~and~\ref{fig:microwave}. }
\end{figure}

As the microwave frequency is scanned we measure the $36d$ state population to reveal the spectrum of excited Stark clusters as shown in Fig.~\ref{fig:microwave}.  Each panel in Fig.~\ref{fig:microwave} shows the microwave spectrum for a particular time delay between the laser excitation pulse and the microwave probe pulse, which is the time during which atoms can interact and exchange energy.  Both pulses have a width of 50~ns. A delay of 0~ns marks the time at which the trailing edge of the laser pulse is aligned with the rising edge of the microwave pulse.  At 7.15~V/cm, the $36d$ state is split into five states; $j=\frac{5}{2},\ |m_j|=\frac{1}{2},\ \frac{3}{2},\ \frac{5}{2}$, and $j=\frac{3}{2},\ |m_j|=\frac{1}{2},\ \frac{3}{2}$.  The range of energy between these five states is slightly smaller than the separation between neighboring manifold clusters.  We therefore expect the microwave spectrum to show five peaks for a single Stark manifold cluster.  The colored bands in Figs.~\ref{fig:starkmap}~and~\ref{fig:microwave} mark the range of microwave resonances associated with a single cluster.  To find the population in each Stark cluster, we integrate the signal in each band.  The resulting energy transport due to dipole-dipole interaction is shown in Fig.~\ref{fig:time}, where the color and dashing of each curve matches that of Figs.~\ref{fig:starkmap}~and~\ref{fig:microwave}.  The error bars represent the statistical uncertainty associated with averaging 80 measurements at each microwave frequency and each time delay.

At an interaction time of $-100$~ns there should be no signal in the $36d$ state since the microwave probe pulse turns off before the Rydberg excitation pulse turns on.  In fact, we see a small amount of signal in the $36d$ state due to the finite turnoff time of the microwave pulse.  As the delay is increased to 0~ns, we see the overall signal rise as more Rydberg atoms are excited, as well as population in neighboring Stark clusters.  Beyond 0~ns, the number of excited atoms remains constant while the interaction continues to spread population to neighboring manifold clusters, although this spread is rather slow, leaving the bulk of the population in the initial Stark cluster.  

\begin{figure}
 \includegraphics{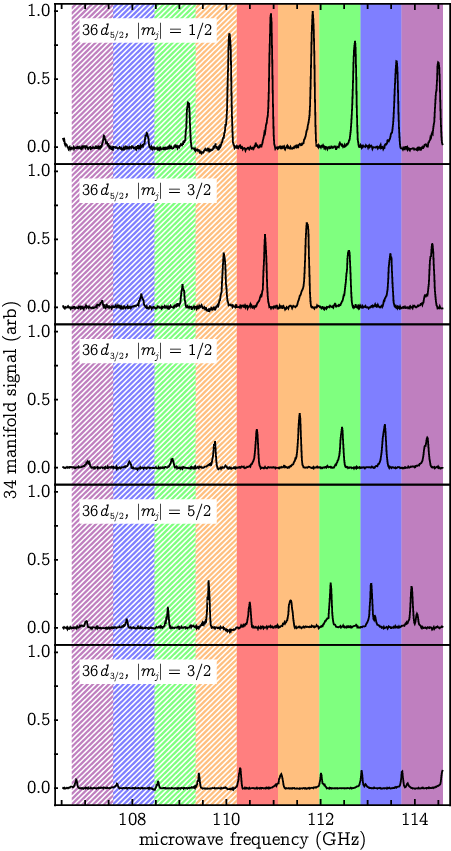}%
\caption{\label{fig:scale}Microwave spectra taken by exciting each of the five $36d$ states individually and driving population down to the manifold clusters below.  This reveals significant variation in the coupling strengths across the manifold.}
\end{figure}

\begin{figure}
 \includegraphics{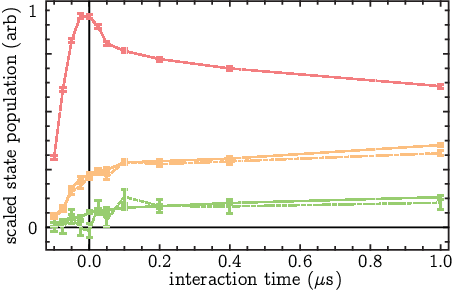}%
\caption{\label{fig:timescaled}Using the measured microwave coupling strengths, we scale the time dependence data for the five central clusters.}
\end{figure}

This integrated signal gives us a glimpse into how energy spreads through our system.  To turn this into a quantitative population distribution measurement, we must account for the variation in coupling strength between the $36d$ states and each manifold cluster.  We measure this coupling strength directly in a separate experiment in which we switch to exciting each $36d$ state individually~\cite{fahey_ExcitationRydbergStates_2011}.  The microwave field is then scanned over the same frequency range as before to drive population down to various Stark clusters.  The results of these measurements are shown in Fig.~\ref{fig:scale}.  Integrating these data across the same color bands, we find that the relative coupling strengths between each manifold cluster and the $36d$ states are (0.019, 0.022, 0.046, 0.116, 0.147, 0.194, 0.151, 0.145, 0.160), corresponding to clusters $-4$ to $4$.  Using these relative coupling strengths, we convert the data in Fig.~\ref{fig:time} into a relative population distribution for each time delay. This evolution of population spread with time is shown in Fig.~\ref{fig:timescaled} for the five central clusters measured.  Although population does spread beyond these five clusters, the data at small microwave frequency, for clusters above the initial cluster, are quite noisy due to the weak coupling strength in this region.

Ideally, we would quench the system into resonance. However, since the dipole-dipole interactions are resonant at a wide range of electric fields, we cannot excite at a non-resonant field. Thus, interactions occur during the 50~ns laser excitation pulse and some of the energy transport visible before 0~ns in Fig.~\ref{fig:timescaled} is possibly due to direct excitation of pair states~\cite{deiglmayr_ObservationDipoleQuadrupoleInteraction_2014}. However, after 0~ns the initial state population continues to decrease while the populations of other clusters increase, a clear signature of dipole-dipole interactions. In fact, Seiler \textit{et al.} determined that resonant dipole-dipole interactions among manifold states were required to model the decay of Rydberg atom population in an electrostatic trap~\cite{seiler_RadiativeCollisionalProcesses_2016}.

\section{Modeling}

In order to model the energy transport in the manifold, an estimate of the Rydberg density is needed. We estimate the density using the simple set of few-body dipole-dipole interactions studied in~\cite{liu_TimeDependenceFewBody_2020}. We tune our 1265~nm laser to excite Rydberg atoms to the 36$p_{3/2,|m_j|=1/2}$ state in an electric field with no resonant interactions. We quench into resonance, and the 36$s$ state is populated via a two-body interaction at 3.29~V/cm or a three-body interaction at 3.52~V/cm. We measure the 36$s$ state population as a function of interaction time over 1~$\mu$s in 10~ns increments, while interleaving the two resonant electric fields. The diameter of the excitation beam is measured using a knife-edge.

We simulate the time evolution of the 36$s$ state fraction by numerically solving the Schr\"odinger equation in a cylindrical volume with the measured diameter. The only free parameters in the simulation are the density of the Rydberg atoms, the dipole blockade radius, and an overall scale factor. The simulations are compared to the data and the simulated density with the smallest chi-squared value is selected as the best fit. Since the two- and three-body dipole-dipole interactions scale differently with density, the fit is constrained and yields a relatively small range of densities. We estimate the Rydberg density to be $(4\pm 2)\times 10^9$~cm$^{-3}$, or an average interatomic spacing of about 6~$\mu$m. 

Using the calibrated density we model the manifold system by numerically solving the Schr\"odinger equation using the two-body Hamiltonian
\begin{equation}
    \sum_{a}\sum_{i}\hat{\sigma}_{aa}^i E_a 
    +  \sum_{a \ne b}\sum_{c \ne d}\sum_{i \ne j} \left(\hat{\sigma}_{ab}^i \hat{\sigma}_{cd}^j +\mathrm{H.c.} \right)\frac{\mu_{ab}\mu_{cd}}{R_{ij}^3},\label{eq:ham}\nonumber
\end{equation}
where $\hat{\sigma}^i_{ab}$ is an operator that takes the $i^{th}$ atom from Stark level $a$ to Stark level $b$ and $R_{ij}$ is the distance between the $i^{th}$ and $j^{th}$ atoms. The transition dipole moments $\mu_{ab}$, which connect Stark level $a$ to $b$, are calculated using the Numerov method~\cite{zimmerman_StarkStructureRydberg_1979}. The first term gives the energy of each level and the second term represents the two-body dipole-dipole interactions. When $c=b$ and $d=a$, the second term yields hopping interactions.

All of our simulations assume that the atoms are motionless and we ignore the angular dependence of the interactions. We randomly place dozens of atoms in a spherical volume at the measured density. Including enough energy levels to faithfully represent the manifold necessarily limits the number of atoms we can include in the Schr\"odinger equation. We iterate over each atom in the volume, select its three nearest neighbors, and then run the simulation with those four atoms. Including at least four atoms allows for the the possibility of ``secondary'' interactions. For example, two pairs of atoms in the initial cluster could exchange energy, resulting in a pair of atoms in other energy clusters that are now able to interact. We repeat this process for multiple volumes and average so that the final results include a few hundred simulations.

With only four atoms in the model, the simulated dynamics will be slower than in the experiment~\cite{spielman_quantum_2024b}. Since we also do not model direct excitation of pair states, we do not expect our simulations to be quantitatively accurate. However, we can look at longer simulated times, when the populations of each energy cluster have reached a steady state.

\begin{figure}
 \includegraphics{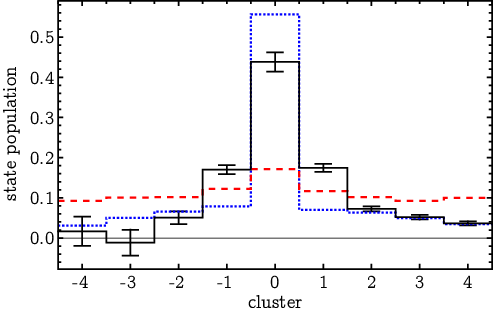}%
\caption{\label{fig:model} Fraction of atoms as a function of energy for the experimental data of Fig.~\ref{fig:microwave} at $t=1$~$\mu$s (black), a simple harmonic ladder (dashed red), and a more complete model of the Stark manifold (dotted blue). The total signal is binned for each cluster and the experimental data has been normalized to the total measured signal across all clusters. The statistical uncertainty in the simulated results is small compared to the experimental uncertainty and is not shown.}
\end{figure}

As a starting point for understanding the spread of energy in our system, we consider a perfectly harmonic ladder of nine energy levels with four atoms initially in the middle level. We set the separation between energy levels to be 0.8~GHz, the mean separation between clusters in the actual experiment. The $\mu_{ab}$ are randomly drawn from a box distribution centered on the mean between-cluster dipole moment. Figure~\ref{fig:model} compares the distribution of energy, averaged over the last five microseconds of a 100-microsecond simulation, to the experimental data. The simulation has equilibrated with a relatively even energy distribution. When we run this model with five or six atoms and more energy levels, we see similar results.

The predictions of the harmonic ladder model disagree strongly with the experimental result. We can try to improve the model by adding an anharmonicity similar to that of the actual energy-level clusters in the Stark map of Fig.~\ref{fig:starkmap}. Moving outward from the initial cluster used in the experiment, the mean energy of each cluster shifts by about 0.1\% of the cluster spacing. We find that the results do not significantly change when we include an experimentally realistic anharmonicity. 

Finally, we create a more realistic model by including clusters of energy levels taken from the calculated Stark energy levels of Fig.~\ref{fig:starkmap}, along with the calculated $\mu_{ab}$. We include nine clusters of four $\vert m_j\vert=\frac{1}{2}$ and $\frac{3}{2}$ energy levels. We include only four-atom states whose detuning from the initial state is less than the typical energy spacing between clusters, resulting in about $125\,000$ states. 

While the improved model still ignores many energy levels with $\vert m_j\vert > \frac{3}{2}$, the simulation results shown in Fig.~\ref{fig:model} are in qualitative agreement with the data. This model shows a similar degree of symmetric energy spreading to more distant clusters, while the bulk of the population remains in the initial cluster. The small anharmonicity should play even less of a role than in the simple model, since the width of each cluster allows for near-resonant interactions even in the mean spacing between clusters has shifted. Thus, our results suggest that the equilibrium energy distribution is due to the structure of the clusters. 

The dynamics are strikingly simple given the complex structure of the energy levels and the couplings $\mu_{ab}$. Some of this simplicity may be attributed to our choice of initial cluster. As seen in Fig.~\ref{fig:starkmap}, the initial cluster is so tightly spaced that the individual energy levels are not resolved. The bandwidth of our excitation laser is broad enough to excite a superposition of these energy levels. With so many possible initial pair states, there are dipole-dipole interactions with small detuning to every final pair of clusters.

Because an excess of population remains in the initially excited cluster after equilibration, it appears that the system has failed to thermalize. However, this assumes that the thermal equilibrium state is, indeed, a more even distribution of energy. In future work, we plan to calculate the thermal state for comparison.

\section{Conclusion}

Our results indicate that the Rydberg Stark manifold offers a promising platform for studying quantum many-body dynamics with a rich and complex energy level structure. Experiments with dipolar energy exchange in atomic systems usually start with a simpler set of energy levels. The many closely-spaced energy levels in the manifold are more similar to solid-state systems, offering prospects for quantum emulation. While the energy levels discussed in this experiment are arranged into nearly harmonic clusters, the addition of a static magnetic field could either broaden those clusters or cause them to overlap. For $|m_j| > \frac{5}{2}$, there are also other nearly harmonic ladders that fall between the ladder studied here. Furthermore, we find that the rich level structure in this work has strong potential to be realized as a synthetic dimension, though the all-to-all coupling present in this system poses a unique challenge for simulating typical lattice hopping.

Interactions beyond two-body energy exchange are also possible in the manifold. In fact, nearly every two-body interaction studied here is somewhat detuned from resonance. Since there are many energy-level spacings available within the clusters, the addition of a third atom can bring the interaction closer to resonance. While these three-body interactions scale as $R^{-6}$, with their smaller detuning we expect that they contribute significantly at higher densities. A study of the density dependence of energy transport in the manifold may be able to detect their effect.

\begin{acknowledgments}
This work was supported by the National Science Foundation under Grants No. 2011583 and No. 2011610, and S.E.S. is supported by the National Science Foundation Graduate Research Fellowship under Grant No. 2334429.

This work used the Delta system at the National Center for Supercomputing Applications through allocation PHY230142 from the Advanced Cyberinfrastructure Coordination Ecosystem: Services \& Support (ACCESS) program, which is supported by National Science Foundation grants \#2138259, \#2138286, \#2138307, \#2137603, and \#2138296.
\end{acknowledgments}

\bibliography{StarkDDtime}

\end{document}